\documentclass[trackchanges]{aastex701}

\usepackage{graphicx}
\usepackage{color}
\usepackage{xcolor}
\usepackage{comment}
\usepackage{orcidlink}

\newcommand{\kms}{km\,s$^{-1}$}

\begin{document}

\title{Oblique kink waves in solar coronal streamers}

\author[orcid=0000-0001-5358-2153, gname=Aleksandr, sname=Rubtsov]{Aleksandr V. Rubtsov}
\affiliation{Key Laboratory of Solar Activity and Space Weather, School of Aerospace, Harbin Institute of Technology, Shenzhen, Guangdong 518055, China}
\email[]{rubtsov@hit.edu.cn}  

\author[orcid=0000-0001-6423-8286, gname=Valery, sname=Nakariakov]{Valery M. Nakariakov}
\affiliation{Centre for Fusion, Space and Astrophysics, Department of Physics, University of Warwick, CV4 7AL, Coventry, UK}
\affiliation{G-LAMP NEXUS Institute/School of Space Research, Kyung Hee University,  Yongin, 17104, Republic of Korea}
\affiliation{Centro de Investigacion en Astronom\'ia, Universidad Bernardo O'Higgins, Avenida Viel 1497, Santiago, Chile}
\email{V.Nakariakov@warwick.ac.uk}

\author[orcid=0000-0002-3278-6250, gname=Dmitri, sname=Klimushkin]{Dmitri Yu. Klimushkin} 
\affiliation{Institute of Solar-Terrestrial Physics SB RAS, Irkutsk, 664033, Russia}
\email{klimush@iszf.irk.ru}

\author[orcid=0000-0002-9514-6402, gname=Ding, sname=Yuan]{Ding Yuan}
\affiliation{Key Laboratory of Solar Activity and Space Weather, School of Aerospace, Harbin Institute of Technology, Shenzhen, Guangdong 518055, China}
\email[show]{yuanding@hit.edu.cn}

\begin{abstract}
Coronal streamer kink waves excited by an impact of coronal mass ejections (CMEs) are routinely used to estimate the coronal Alfv\'en speed. These estimates generally assume wave propagation parallel to the equilibrium magnetic field directed radially outward to the Sun, in the plane of the sky. However, the geometry of CME impacts naturally allows for oblique propagation. We investigate the properties of obliquely propagating kink waves guided by a plasma slab with a central current sheet representing a coronal streamer stalk. We consider both cold and warm streamer plasmas and analyse influence of propagation obliquity on wave dispersion, eigenfunctions, and seismological diagnostics. We find that increasing the propagation obliquity reduces the phase speed owing to both projection effects and the intrinsic dependence of the phase speed on the propagation angle. The wave structure also undergoes a gradual transition from a body mode to a surface mode, leading to stronger perturbations at the slab boundaries and enhanced wave localization. Consequently, for a given energy deposited by a CME impact, oblique kink waves are expected to produce larger observable amplitudes than field-aligned waves. Applying the developed theory to streamer waves reported previously, we infer Alfv\'en speeds in the range of 50--500~\kms. We show that neglecting wave propagation obliquity may underestimate the Alfv\'en speed by up to a factor of 1.5. These results demonstrate that wave propagation obliquity should be taken into account in coronal seismology of streamer waves and that stereoscopic observations will be required to constrain the true propagation direction.
\end{abstract}


\section{Introduction}

Coronal streamers attract attention as possible sources of slow solar wind \citep[see, e.g.,][for reviews]{1992SSRv...61..393K, 2004AdSpR..33..681O, 2016SSRv..201...55A, 2020ApJ...893...57M}. 
Streamers appear as bright, helmet-shaped structures extending outward from the Sun, as seen in white-light observations. They are associated with closed magnetic field regions that connect opposite-polarity areas at the solar surface, forming a large-scale loop system. Above that loop lies a cusp-shaped structure, beyond which the magnetic field is believed to open into the heliosphere, forming a radial stalk aligned away from the Sun. Streamers are observable for long periods of time, covering several solar rotations \citep[e.g.,][]{2008ApJ...680.1532Z}. In white-light, the stalk appears as a thin, elongated, ray-like feature. Typically, the stalk is rooted in the cusp at a distance of about 1--2 solar radii, $R_\odot$, from the surface, extending up to tens of $R_\odot$ into the heliosphere. The width is approximately 0.05--0.2~$R_\odot$ near the cusp, increasing to 0.1--0.3~$R_\odot$ at 5--10~$R_\odot$. Near the root, the plasma density is estimated to be about 10$^6$--10$^7$~cm$^{-3}$, and decreases with height. The plasma temperature is about 1--2 million degrees. Typical values of the magnetic field are estimated as 1–-2~G at lower heights. The plasma parameter $\beta$ is expected to increase with height, becoming greater than unity \citep[e.g.,][]{2015SoPh..290.2043A, 2023ApJS..265...47H}.

Coronal streamers are highly dynamic structures. Observations show a wide range of processes that involve magnetic reconnection, magnetohydrodynamic (MHD) waves, plasma flows, and instabilities \citep[e.g.,][]{1973Natur.246..414K, 1997ApJ...484..472S, 2004AdSpR..33..681O, 2013ChSBu..58.1599C, 2026ApJ...999L..41Z}. Among the extensively studied dynamic phenomena are kink oscillations of streamer stalks. Kink oscillations appear as alternate transverse displacements of an elongated plasma non-uniformity, see, for example, kink oscillations of coronal loops \citep[e.g.,][]{2021SSRv..217...73N}. The kink oscillations of streamers, known as \lq\lq streamer waves\rq\rq, were first detected as outward propagating transverse displacements of a streamer stalk by \citet{2010ApJ...714..644C, 2011SoPh..272..119F} in white light  with the Large Angle and Spectrometric Coronagraph (LASCO) on the Solar and Heliospheric Observatory (SOHO). The value of the phase speed projected on the plane of the sky exceeded the sound speed. Follow-up case studies include \citep{2011ApJ...728..147C, 2011SoPh..272..119F, 2013ApJ...766...55K, 2016MNRAS.463.1409S}. The latter work revealed acceleration of the kink wave with height. 
\citet{2020ApJ...893...78D} studied 22 streamer kink wave events detected with the white-light coronagraph COR2 on the Solar TErrestrial RElations Observatory (STEREO), and LASCO  C2 and C3, which allowed the authors to determine typical parameters of this phenomenon. Streamer waves were found to propagate at phase speeds of 360--740~\kms. Typical displacement amplitudes are 0.05--0.5~$R_\odot$, i.e., greater than the width of the stalk. Typical wavelengths are 3--7~$R_\odot$ and oscillation periods are a few hours. All of the detected streamer waves decay in about a wavelength.  

Recently, \citet{2026Univ...12...89S} observed two consecutive kink waves propagating along the same streamer stalk, from two different lines-of-sight with COR2 on STEREO A and LASCO C2 and C3. The waves had wavelengths of about 4.0~$R_\odot$ and 6.2~$R_\odot$, and periods of 2.7 and 2.5~hours, respectively. Both waves were seen to decelerate with distance from the Sun. The amplitude of the first wave decreased with distance from the Sun, while the second wave exhibited an increase in amplitude. 

Although streamer kink waves have been also suggested to be caused by the Kelvin--Helmholtz instability \citep{2013ApJ...774..141F}, the most popular interpretation associates them with fast magnetoacoustic modes of the plasma slab forming the streamer stalk \citep[e.g.,][]{2019ApJ...883..152D,2024A&A...682A.168S}. Streamer waves are excited by the impact of a coronal mass ejection (CME). The poor correlation between the wave speed and the speed of the corresponding source CME \citep{2020ApJ...893...78D} indicates that the wave speed is determined mainly by the physical properties of the streamer rather than by the triggering CME. This finding supports the interpretation of streamer waves as eigenmodes of the waveguiding plasma non-uniformity.

Theoretically, the streamer stalk is usually modelled as a slab of enhanced plasma density, which surrounds a current sheet \citep[e.g.,][]{2024A&A...682A.168S, 2025A&A...701A.166S}. The width of the slab is much larger than the true thickness of the current sheet. The guiding magnetic field is usually neglected, i.e., the unperturbed magnetic field is assumed to have only radial components directed oppositely on either side of the current sheet. The Alfv\'en speed inside the slab is lower than outside it.
In such a structure, fast waves are locally oblique with respect to the plane of the current sheet and propagate globally along the slab because of total internal reflection from the slab boundaries. Inside the slab, the perturbations are oscillatory, whereas outside the slab they decrease exponentially with distance from the boundary. Such modes are known as trapped body modes. In terms of ideal MHD, when the current sheet is not subject to Ohmic diffusion and reconnection, the modes of the stalk have properties similar to fast waves guided by a slab penetrated by a unidirectional magnetic field \citep[see, e.g.,][]{1982SoPh...76..239E, 1995SoPh..159..399N, 2015ApJ...814...60Y, 2015ApJ...801...23L, 2022MNRAS.515.4055G}.

In the case of waves propagating along the anti-parallel magnetic field, i.e., radially away from the Sun, the phase speed of the kink mode lies between the Alfv\'en speeds inside and outside the slab. In the long wavelength limit, when the wavelength is much longer than the width of the slab, the phase speed of the kink mode tends to the external Alfv\'en speed. This makes streamer waves a promising probe of the magnetic field in the streamer stalk through the method of coronal seismology, e.g., \citet{2011ApJ...728..147C, 2011SoPh..272..119F, 2016MNRAS.463.1409S, 2020ApJ...893...78D}; see, also, \citet {2024RvMPP...8...19N} for a recent comprehensive review of the method. 
Alternative techniques for coronal magnetic field estimation  based on the extrapolation of the photospheric field and radio emission have serious limitations \citep[e.g.,][]{2009ApJ...696.1780D, 2018SSRv..214...99Y, 2025A&A...697L...9K}.  
Furthermore, kink waves guided by streamers can be used to probe the acceleration of the nascent solar wind. This technique is complementary to, or perhaps an alternative to, the currently used blob-tracking method \citep[e.g.,][]{2009ApJ...694.1471S}. 

The dependence of the oscillation period on the wavelength appears to be rather scattered \citep{2011ApJ...728..147C, 2011SoPh..272..119F, 2016MNRAS.463.1409S, 2020ApJ...893...78D}, which may indicate a wide spread in the values of the Alfv\'en speed in and around the stalk. However, the impact of a CME may excite kink waves guided by the stalk slab obliquely, i.e., at an angle to the radial direction away from the Sun. Hence, non-parallel kink modes of a slab, propagating at an angle to both the radially directed magnetic field and the direction of the transverse plasma non-uniformity, are of particular interest. The properties of oblique kink waves can differ significantly from those in the parallel-propagation limit \citep[e.g.,][]{2007SoPh..246..213A, 2023MNRAS.518L..57L, 2026A&A...707A.273L}. The phase speed of such waves projected onto the plane of the sky may differ significantly from the actual phase speed, which needs to be taken into account in seismological estimations. In addition, for large perpendicular wave numbers, the kink wave inside the slab changes its nature: the body mode becomes a surface mode.
Furthermore, the detection of kink perturbations propagating obliquely to the line-of-sight (LoS) could be significantly affected by the LoS integration effect in the optically thin emission regime typical of the corona \citep[see, e.g.,][]{2003A&A...397..765C}.

The purpose of this paper is to determine the phase speeds of kink waves guided by a plasma slab surrounding a current sheet, with the magnetic field being anti-parallel on either side of the sheet, for waves propagating obliquely. The main novel element of our study is its specific focus on streamer waves in the context of seismological estimations. In addition, we estimate the perturbation localisation distance outside the slab, which is important for wave detectability.

The paper is organised as follows. In Section~\ref{Sec:mod}, we introduce the model, describe the simplifying assumptions, and derive the dispersion relations. 
Section~\ref{Sec:res} presents the dependence of the radial phase speed, the structure of the perturbation inside the slab, and the extent of the localisation of the mode outside the slab on wave obliquity. 
Section~\ref{Sec:seis} discusses the seismological implications of the results obtained. 
Finally, Section~\ref{Sec:con} summarises our conclusions.

\section{Model and dispersion relations}
\label{Sec:mod}

\begin{figure}
    \centering
    \includegraphics[width=\linewidth]{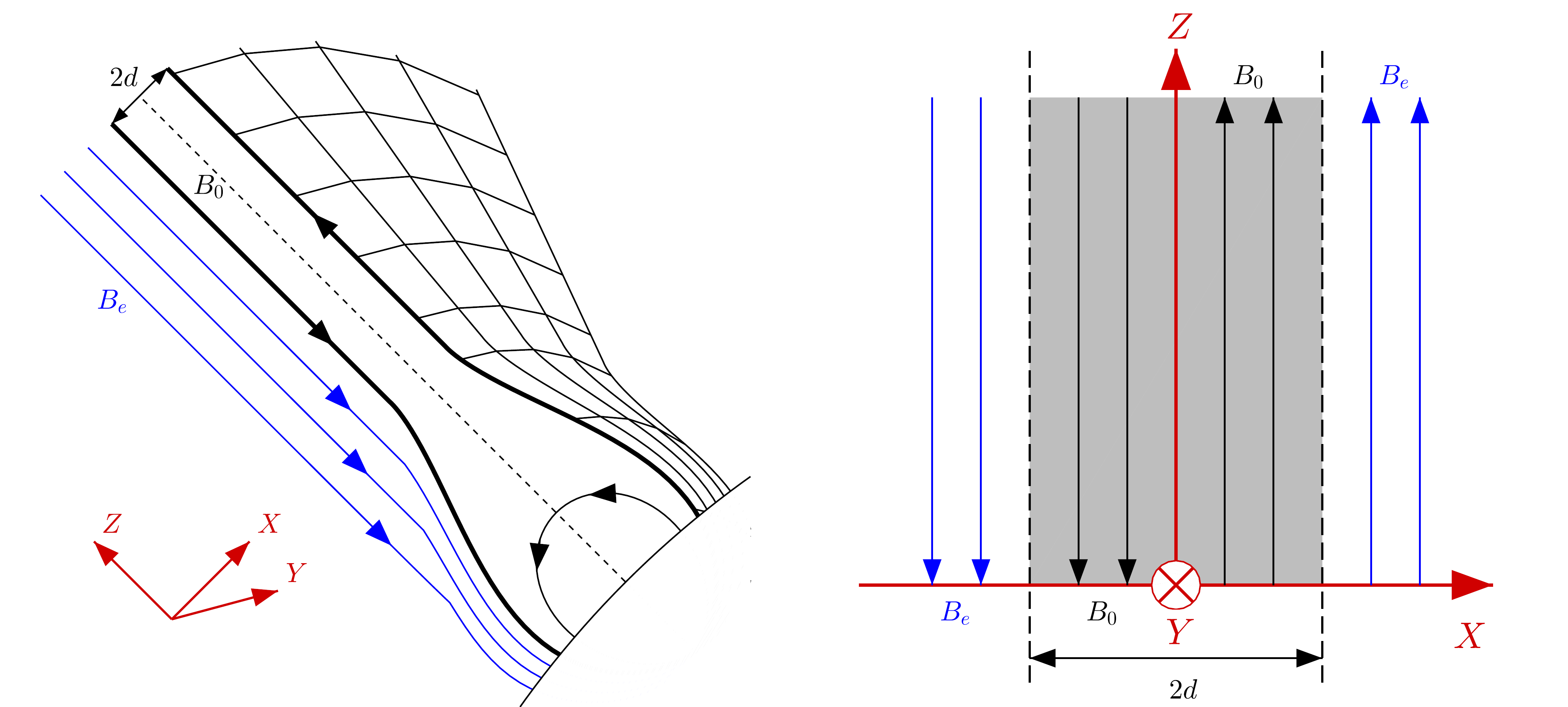}
    \caption{A schematic illustration of a coronal streamer magnetic field geometry with a 3D slab as the stalk (left). The model in the plane $x$-$z$ (right). The grey region is the plasma slab surrounding a current sheet situated at $x=0$. }
    \label{fig:Sketch}
\end{figure}

We consider a streamer stalk as a three-dimensional magnetic slab with the equilibrium plasma density $\rho_\mathrm{0}$ confined to a region $|x| < d$, outside of which the density is $\rho_\mathrm{e}$ (see Figure~\ref{fig:Sketch}). In the following, the indices  $0$ and $\mathrm{e}$ denote physical quantities inside and outside the slab, respectively. 
The slab surrounds a current sheet, with the magnetic field being anti-parallel on either side of the sheet, directed everywhere in the $z$-direction. The width of the current sheet is assumed to be much smaller than that of the slab and is therefore assumed to be infinitesimally thin.
The absolute values of the external and internal fields are $|B_e|$ and $|B_0|$, respectively. In equilibrium, the total pressures inside and outside the slab are equal to each other. 
The internal and extremal Alfvén speeds are $V_\mathrm{A_{0,e}}$, the sound speeds are $V_\mathrm{s{0,e}}$, and the tube (or cusp) speeds are $V_\mathrm{t{0,e}}^2 = V_\mathrm{A{0,e}}^2 V_\mathrm{s{0,e}}^2 / (V_\mathrm{A{0,e}}^2 + V_\mathrm{s{0,e}}^2)$, respectively. 
In the equilibrium, the density contrast $\rho_\mathrm{0} / \rho_\mathrm{e}$ is related to the characteristic speeds as
\begin{equation} \label{eq:denscontr}
\frac{\rho_\mathrm{0}}{\rho_\mathrm{e}} = \frac{V_\mathrm{se}^2 + V_\mathrm{Ae}^2 \gamma/2}{V_\mathrm{s0}^2 + V_\mathrm{A0}^2 \gamma/2},
\end{equation}
where a constant $\gamma$ is the adiabatic index, i.e., the ratio of specific heats.

Consider linear perturbations of the equilibrium, propagating obliquely to the equilibrium field, i.e., with a component of the wave vector off the plane of the sketch in Figure~\ref{fig:Sketch}. The perturbations are   described by the set of ideal MHD equations. The dependencies on time and the coordinates $y$ and $z$ are considered harmonic,
$\propto \exp (i\omega t - ik_yy - ik_zz)$, where $\omega$, $k_y$ and $k_z$ are the frequency, and the $y$ and $z$ components of the wave vector, $\mathbf{k}$, respectively. The structure of the perturbations across the slab is determined by the non-uniformity. 

Generalising the work of \citet{1982SoPh...76..239E}, who considered a two-dimensional problem with the wave vector $\mathbf{k}$ lying in the $x$--$z$ plane, and following, e.g., \citet{2023MNRAS.518L..57L, 2026A&A...707A.273L}, we consider wave propagation oblique to the magnetic field direction $\mathbf{\hat{z}}$, i.e. $k_y \neq 0$. 
Inside and outside the slab, the disturbances normal to the boundary of the slab are described by the equations
\begin{equation} \label{eq:diffeq}
\frac{\text{d}^2 v_x}{\text{d}x^2} - \Big(m_\mathrm{0,e}^2 + k_y^2\Big) v_x = 0,
\end{equation}
where
\[m_\mathrm{0,e}^2 = \frac{(V_\mathrm{A0,e}^2 k_z^2 - \omega^2) (V_\mathrm{s0,e}^2 k_z^2 - \omega^2)}{(V_\mathrm{A0,e}^2 + V_\mathrm{s0,e}^2) (V_\mathrm{t0,e}^2 k_z^2 - \omega^2)}.\]
In the following, it is convenient to introduce quantities $M_\mathrm{0,e} = \sqrt{m_\mathrm{0,e}^2 + k_y^2}$ that are effective wave numbers of the perturbations across the slab. In this study, we focus on wave disturbances that are evanescent in $|x| > d$, i.e., trapped modes, with $M_\mathrm{e}^2 > 0$. 
Thus, the real quantity $M_\mathrm{e}^{-1}$ is the characteristic distance of the exponential decay of the perturbations outside the slab, which can be considered as the mode localisation distance. Solving Eqs.~(\ref{eq:diffeq}) inside the slab, we select either oscillatory ($M_\mathrm{0}^2 < 0$, for body modes) or hyperbolic ($M_\mathrm{0}^2 > 0$, for surface modes) solutions. We then match the internal and external solutions at the slab boundaries, $x = \pm d$, by imposing continuity of the perpendicular displacement and total pressure perturbation. In the static case, without stationary flows, the continuity of the transverse displacement is equivalent to the continuity of the transverse velocity $v_x$ \citep{1995SoPh..159..213N}.
This standard procedure yields the following dispersion relations:
\begin{equation} \label{eq:dispeq}
\rho_e (V_\mathrm{Ae}^2k_z^2 - \omega^2) M_\mathrm{0} \left\{ \begin{array}{cc}
     \coth{M_\mathrm{0} d}  \\
     \tanh{M_\mathrm{0} d} 
\end{array}   \right\} + \rho_0 (V_\mathrm{A0}^2 k_z^2 - \omega^2) M_\mathrm{e} = 0,
\end{equation}
where the choice between $\coth$ or $\tanh$ corresponds to the kink and sausage modes, respectively. This is a general equation for which solutions depend on the hierarchy of characteristic speeds of internal and external media. 

In the following, we set the external Alfvén speed to be higher than the internal $V_\mathrm{Ae} > V_\mathrm{A0}$, and the plasma beta outside the slab, $\beta_\mathrm{e}$ to be zero ($V_\mathrm{se} = V_\mathrm{te} = 0$). Inside the slab, we consider two models: (1) the plasma $\beta_0$ is zero (i.e., $V_\mathrm{s{0}} = V_\mathrm{t{0}} = 0$); and (2) the plasma pressure is higher than the magnetic pressure, $\beta_0 > 1$, i.e., $V_\mathrm{Ae} > V_\mathrm{s0} > V_\mathrm{A0} > V_\mathrm{t0}$. The adiabatic index is $\gamma = 5/3$. In the $\beta_0 = 0$ model, the right hand side of Eq.~(\ref{eq:denscontr}) reduces to the ratio of Alfvén speeds squared. We set $V_\mathrm{Ae} = 2 V_\mathrm{A0}$, which means $\rho_0 / \rho_e = 4$. In model 2, with $\beta_0 > 1$, we maintain the density contrast $\rho_0 / \rho_e \approx 4$ by establishing $V_\mathrm{Ae} = 3.85 V_\mathrm{A0}$ and $V_\mathrm{s0} = 1.5 V_\mathrm{A0}$.

Since the streamer waves of interest in this study are kink oscillations of the streamer stalk, we restrict our attention to kink mode solutions of Eq.~(\ref{eq:dispeq}). The roots of transcendental algebraic equation (\ref{eq:dispeq}) were found numerically using the open source software Scilab (2025.0.0). In the following, it is convenient to use dimensionless wave numbers, e.g., $k_zd$, $k_yd$, etc., normalised to the half-width of the slab. The case $k_z d = 1$ corresponds to the wavelength in that direction being approximately three times the width of the slab, which corresponds to observations.

\section{Effects of wave obliquity}
\label{Sec:res}
\subsection{Radial phase speeds}
\label{sec:phas}

\begin{figure}
    \centering
    \includegraphics[width=\linewidth]{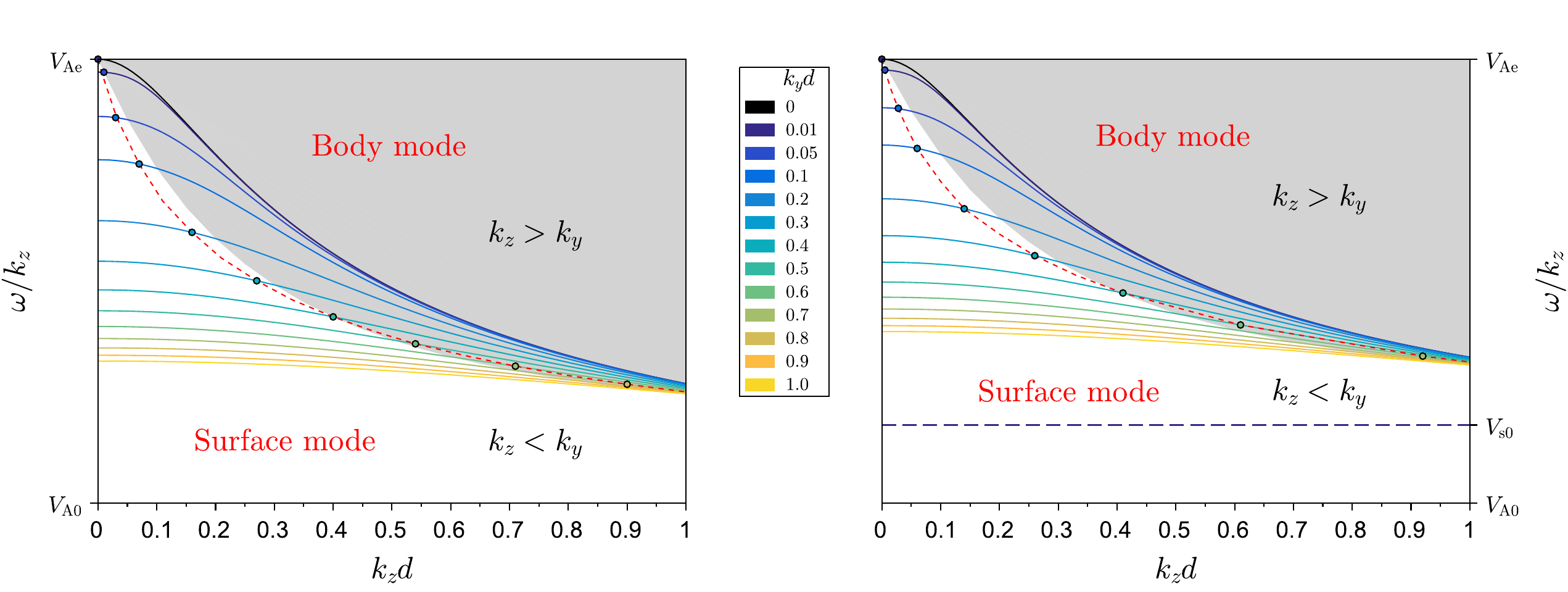}
    \caption{The radial phase speed $\omega/k_z$ of the kink mode of a streamer stalk as a function of the normalised radial wave number $k_z d$ for a set of $k_y d$ shown in the legend. Here $\rho_0 / \rho_e = 4$ and $\beta_\mathrm{e} = 0$. In the left panel, $V_\mathrm{Ae} = 2 V_\mathrm{A0}$,  $\beta_0 = 0$; and in the right panel $V_\mathrm{Ae} = 3.85 V_\mathrm{A0}$,  $\beta_0 = 2.7$ ($V_\mathrm{s0} = 1.5 V_\mathrm{A0}$). The shaded area indicates the $k_z > k_y$ region. The red dashed line with circles denotes the boundary between the body mode and surface mode regimes, $M_0 = 0$.
    }
    \label{fig:KinkPhaseSpeed}
\end{figure}
\begin{figure}
    \centering
    \includegraphics[width=0.5\linewidth]{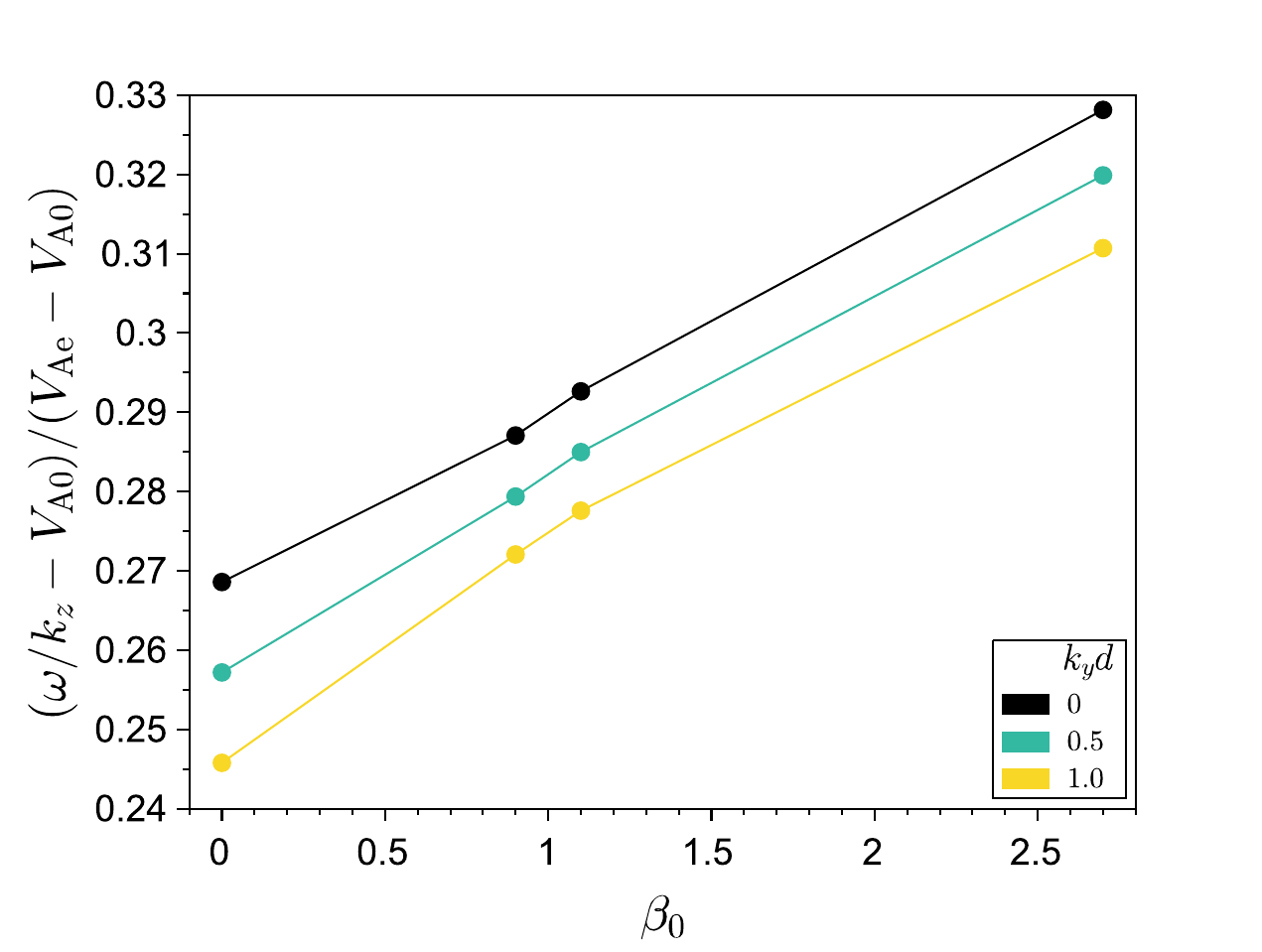}
    \caption{Relative radial phase speed $\omega/k_z$ for $k_zd = 1$ and $k_yd = 0, 0.5$ and 1 against $\beta_0$.}
    \label{fig:beta}
\end{figure}

Consider the radial component of the kink-mode phase velocity $\omega/k_z$ that, in our model, is directed outward from the Sun, along the unperturbed magnetic field. 
Figure~\ref{fig:KinkPhaseSpeed} shows the dependence of the radial phase speed on the radial wave number normalised to the slab half-width, $k_z d$, for different normalised wave numbers in the direction perpendicular to the equilibrium field but parallel to the current sheet and the slab boundaries, $k_y d$. We consider separately the cases of $\beta_0 = 0$ (left panel) and $\beta_0 > 1$  (right panel).

In the $\beta_0 = 0$ case, the radial phase speed is situated between the external and internal Alfvén speeds (Figure~\ref{fig:KinkPhaseSpeed}, left), whereas in the $\beta_0 > 1$ case it lies between the external Alfvén speed and the internal sound speed (Figure~\ref{fig:KinkPhaseSpeed}, right). 
The dispersion curves corresponding to $k_y = 0$ are similar to those obtained by \citet{1982SoPh...76..239E} for parallel propagation.
Increasing $k_y$, i.e. increasing the angle $\theta$ between the wave vector and the equilibrium magnetic field and, hence, the radial direction, reduces the radial component of the phase speed.

According to Figure~\ref{fig:KinkPhaseSpeed}, dispersion of kink waves depends weakly on the specific value of the plasma parameter $\beta_0$ inside the slab. Figure~\ref{fig:beta} shows this dependence for the values of $\beta_0$ situated between the values 0 and 2.7 used in Figure~\ref{fig:KinkPhaseSpeed} for waves with $k_zd = 1$. One can see a smooth increase in the difference between the radial phase speed and the internal Alfv\'en speed with the increase in $\beta_0$.

Furthermore, increasing $k_y d$ leads to the transition from the body-mode regime to the surface-mode regime, as demonstrated, for example, by \citet{2007SoPh..246..213A} in the case $k_z d \ll 1$. The threshold of this transition is determined by the condition $M_0 = 0$ along a dispersion curve, and is indicated in Figure~\ref{fig:KinkPhaseSpeed}. This effect is discussed in more detail in Section~\ref{Sec:stru}.
Waves with low obliquity, i.e., with large values of $k_z/k_y$, are body modes, whereas highly oblique waves, characterised by small values of $k_z/k_y$, are surface modes.
The increase in plasma pressure inside the slab, $\beta_0 > 1$ shifts the mode transition threshold towards smaller obliquity angles $\theta = \arctan({k_y/k_z})$. The increase in the $V_\mathrm{Ae}/V_\mathrm{A0}$ ratio, which keeps the density contrast, compensates for this shift. Thus, the only highly oblique waves are surface modes.

\subsection{Structure of the perturbation inside the slab}
\label{Sec:stru}

\begin{figure}
    \centering
    \includegraphics[width=1\linewidth]{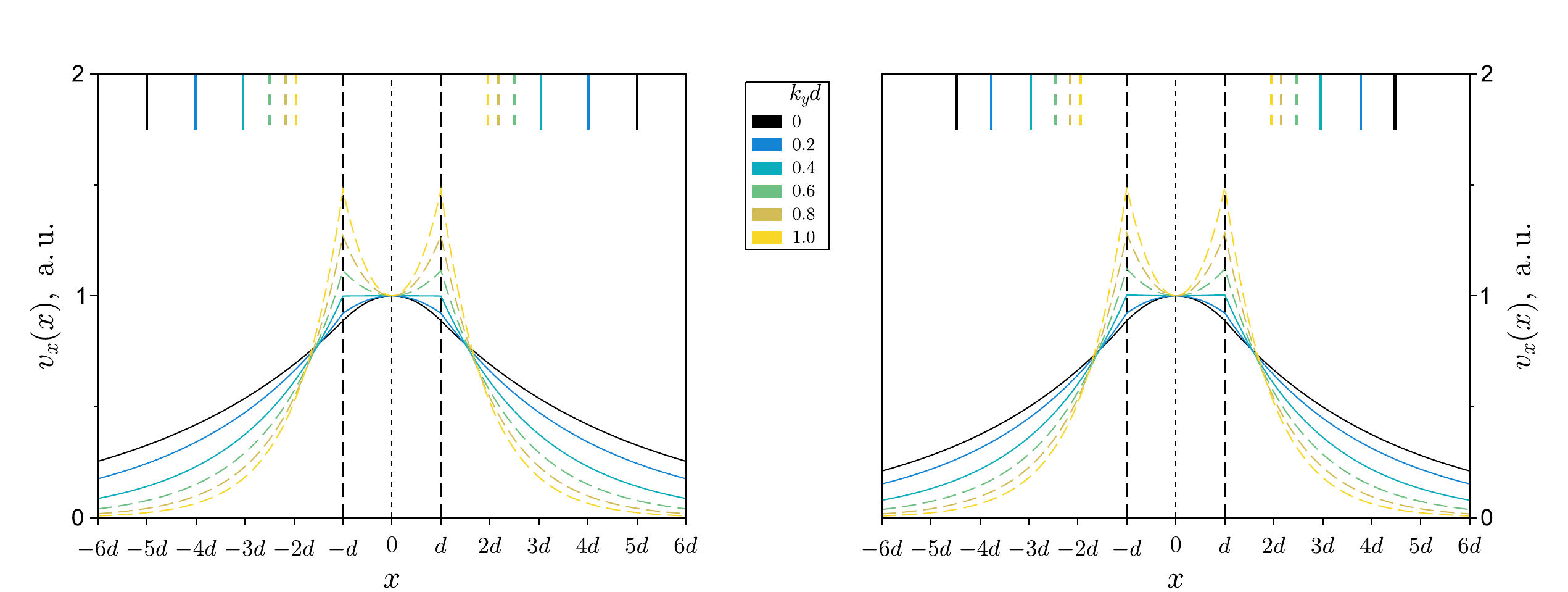}
    \caption{The perpendicular plasma velocity across the slab, $v_x$, as a function of $x$, for the zero-$\beta$ (left) and $\beta_0 > 1$ (right) models. Here $k_z d = 0.4$, and $k_y d$ values are shown in the legend and denoted by different colours. The solid and dashed colour lines correspond to body and surface modes, respectively. All solutions are normalised to the value of the perpendicular velocity at the centre of the slab. The colour vertical marks at the top denote the mode localisation distance outside the slab $1/M_e d$. Vertical dashed lines are the slab boundaries, and the vertical dotted line is the current sheet location.}
    \label{Fig:transvx}
\end{figure}

In the case of parallel propagation, i.e, when the waves propagate radially outward to the Sun, along the field, inside the slab, the waves have a body structure. Oblique waves change their internal structure from body to surface at a certain angle, see Section~\ref{sec:phas}.
Figure~\ref{Fig:transvx} shows the variation in the distribution of the perpendicular component of the velocity perturbation with wave obliquity. 

The key difference between the surface and body kink modes is the location of the maximum perpendicular velocity. For body modes, the maximum velocity is at the centre of the slab, around the current sheet. For surface modes, the perturbation is localised near the boundary of the slab. In addition, in the combination of the parameters considered, outside the slab, the surface modes are localised closer to the slab boundary. This effect is discussed in more detail in Section~\ref{Sec:loc}.

\begin{figure}
    \centering
    \includegraphics[width=0.5\linewidth]{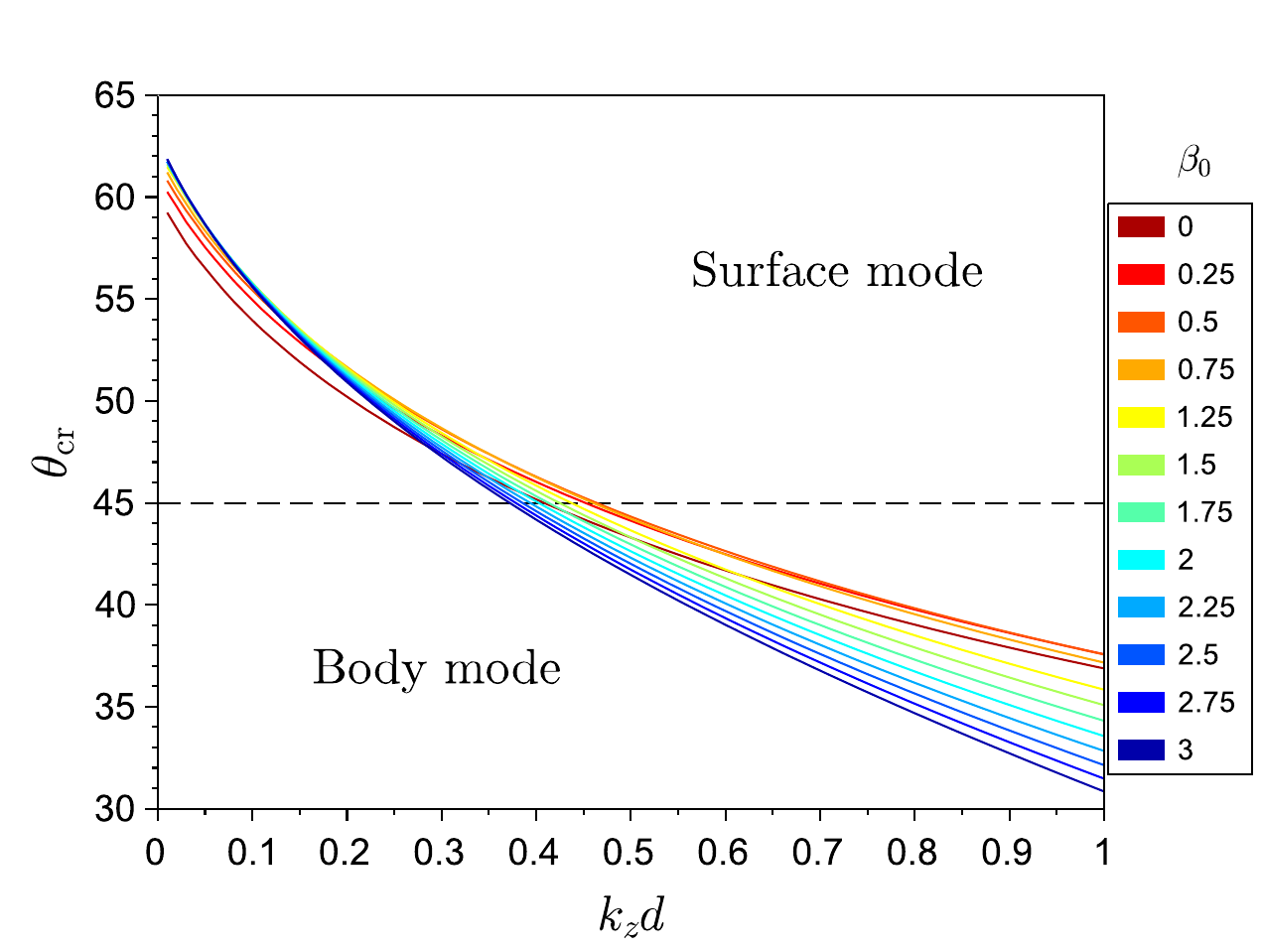}
    \caption{The critical angle from the radial direction for the body--surface mode transition ($M_0 = 0$) as a function of the parallel wave number normalised to the half-width of the slab. Different curves correspond to different values of the internal parameter $\beta_0$ ranging from 0 to 3. The density contrast $\rho_0/\rho_e = 4$. Kink wave propagation angles above each curve correspond to the surface mode, while angles under each line correspond to the body mode. For reference, the horizontal dashed line shows the angle of 45$^\circ$.}
    \label{fig:crit}
\end{figure}

The angle $\theta_\mathrm{cr}$ at which the transition between the body and surface mode regimes inside the slab occurs is determined by the condition $M_0 = 0$. Using this condition, we obtain
\begin{equation}
\label{Eq:loc}
\theta_\mathrm{cr} = \arctan \left(\frac{k_y}{k_z}\right) = \arctan\left(\sqrt{-\frac{m_0^2}{k_z^2}}\right) = \arctan{\sqrt{-\frac{(V_{\mathrm{A0}}^2 - \omega^2 / k_z^2)(V_{\mathrm{s0}}^2 - \omega^2 / k_z^2)}{(V_{\mathrm{A0}}^2 + V_{\mathrm{s0}}^2)(V_{\mathrm{t0}}^2 - \omega^2 / k_z^2)}}}.
\end{equation}
As the radial phase speed depends on the wavelength, so is the critical angle, see Figure~\ref{fig:crit}. The decrease in wavelength causes a decrease in the critical angle. For an important case $k_zd = 1$ corresponding to observations, this regime change occurs at an angle about $35^\circ$ for a wide range of $\beta_0$.

\subsection{Mode localisation distance}
\label{Sec:loc}

An important parameter of a guided wave is the localisation, or e-folding distance, of the perturbation outside the waveguide. 
In our model, the characteristic distance of the mode localisation is $M_e^{-1}$. Under the assumption of $\beta_\mathrm{e} = 0$, the mode localisation distance is estimated as 
\begin{equation}
\label{Eq:loc}
\frac{1}{M_e} = \frac{1}{\sqrt{k_z^2 + k_y^2 - \omega^2/V_\mathrm{Ae}^2}}.
\end{equation}
Hence, the mode localisation distance is determined by the dependence of $\omega$ on the wavelength and the wave obliquity angle $\theta$ determined by the values of $k_y$ and $k_z$. 

\begin{figure}
    \centering
    \includegraphics[width=\linewidth]{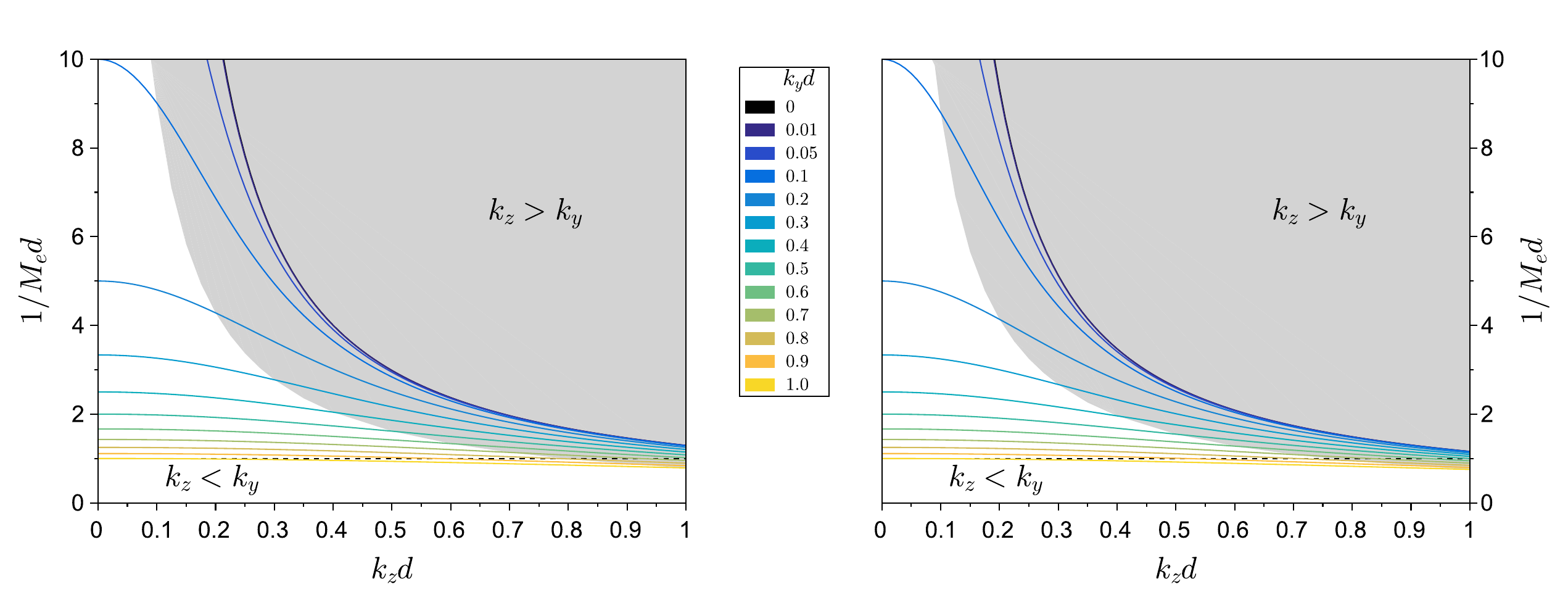}
    \caption{Kink mode localisation outside the slab $1/M_e d$ as a function of $k_z d$ for a set of $k_y d$ shown in the legend for the zero-$\beta$ (left) and $\beta_0 > 1$ (right) models.  The horizontal dashed line denotes $1/M_e d = 1$.}
    \label{fig:invME}
\end{figure}

Introducing the normalisation to the half-width of the slab, $1/M_e d$, we calculate the mode localisation distance as a function of $k_z d$ for a set of values $k_y d$ using solutions to dispersion relation (\ref{eq:dispeq}). The results are shown in Figure~\ref{fig:invME}. The large extension of the mode outside the slab, $1/M_e d \gg 1$, corresponds to longer wavelengths, i.e., when either $k_z d$ or $k_y d$, or both are less than 0.1. The mode localisation distance rapidly decreases for any $k_z d \rightarrow 1$ or $k_y d \rightarrow 1$. For $k_y d \approx 1$ the mode localisation distance is always about or less than $d$, i.e., $1/M_e d \approx 1$. Solutions with lower $k_y d$  cross this threshold at higher $k_z d$, while the localisation distance of kink waves with $k_y d > 0.5$ is less than $2d$. Less oblique waves, with $k_z > k_y$, have larger localisation distances.

\subsection{Friedrichs diagrams}
\label{Sec:phsp}

\begin{figure}
    \centering
    \includegraphics[width=\linewidth]{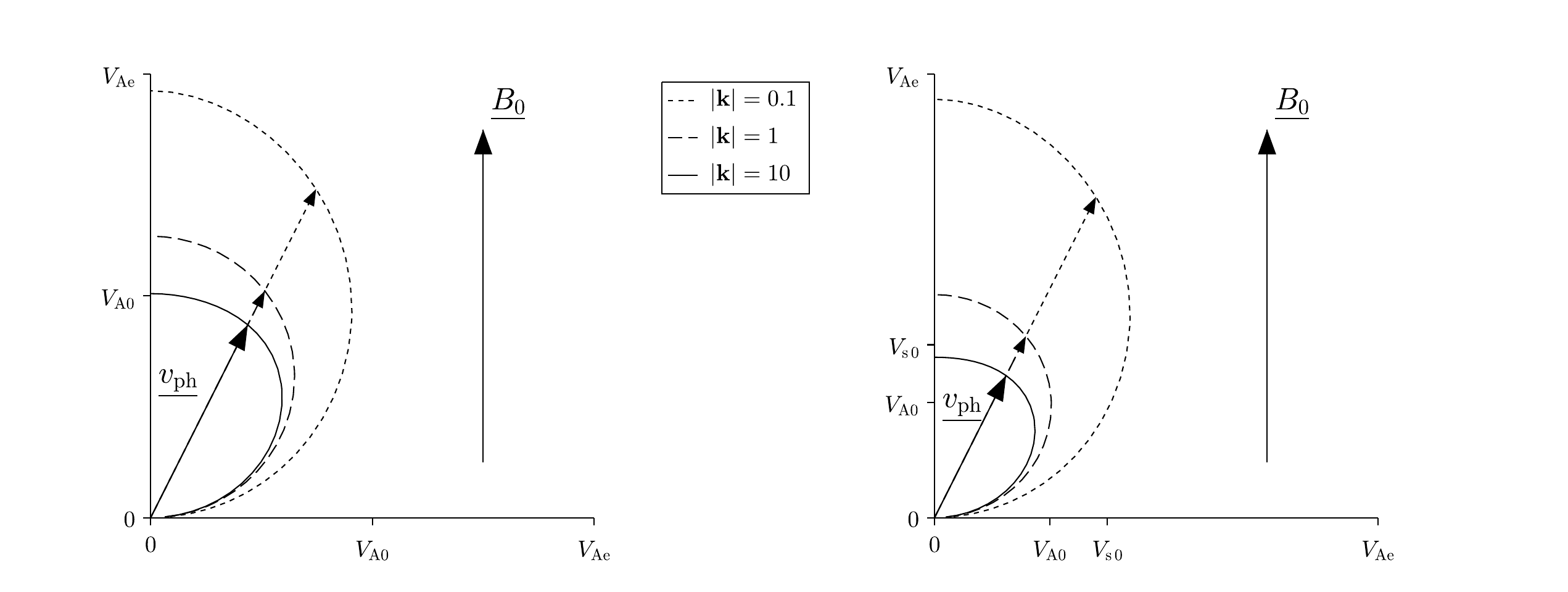}
    \caption{Phase speed $v_\mathrm{ph} = \omega/|\mathbf{k}|$ dependence on obliquity angle for constant $|\mathbf{k}| = \sqrt{k_y^2+k_z^2}$ equals 10 (solid line), 1 (dashed line), and 0.1 (dotted line) for the zero-$\beta$ (left) and $\beta_0 > 1$ (right) models. The equilibrium magnetic field vector is shown by a separate arrow.}
    \label{fig:FriedDiagr}
\end{figure}

A convenient way to demonstrate the anisotropy of the wave propagation, i.e., the effect of obliquity, is the use of Friedrichs diagrams. Polar plots showing the dependence of the phase speed of a guided kink wave on the angle between the wave vector and the direction of the unperturbed field are presented in Figure~\ref{fig:FriedDiagr}. The phase speed decreases with the decrease in the wavelength, and also with the increase in the obliquity. In the case of the highly oblique propagation, the phase speed of kink wave is much lower than both the internal and external Alfv\'en speeds in both $\beta = 0$ and $\beta > 1$ cases. The phase speed approaches zero when the wave vector is perpendicular to the equilibrium magnetic field. This behaviour has already been pointed out in previous findings, see, e.g., \citet{2026A&A...707A.273L}.

\section{Alfv\'en speed estimation}
\label{Sec:seis}

Section~\ref{Sec:res} demonstrates that the phase speed of kink waves is highly sensitive to wave obliquity, which may significantly affect seismological estimates based on measurements of the phase speed projected on the plane of the sky. In the following, we apply the results of  Section~\ref{sec:phas} to the set of streamer waves observed by STEREO/COR2 and SOHO/LASCO coronagraphs \citep{2020ApJ...893...78D}, and evaluate seismological inversions of the Alfv\'en speed.

The data set contains oscillation periods $P$ and wavelength $\lambda$ of 22 streamer waves. The wavelength is projected in the plane of the sky. Assuming that the equilibrium magnetic field is perpendicular to the line-of-sight too, we consider the observed wavelength as the radial component of the full wavelength, that is $\lambda = \lambda_z = 2 \pi /k_z$ in our notation. The ratio of wavelength and period, $\lambda / P$, gives us the phase speed projected in the plane of the sky. In the following, we chose the slab half-width $d = 0.2 R_\odot$ for all waves as an average value at 5$R_\odot$ from the Sun. In \citet{2020ApJ...893...78D}, the measurement error for the period is $\pm$30 minutes, while the wavelength was provided without measurement errors. 

\begin{figure}
    \centering
    \includegraphics[width=1\linewidth]{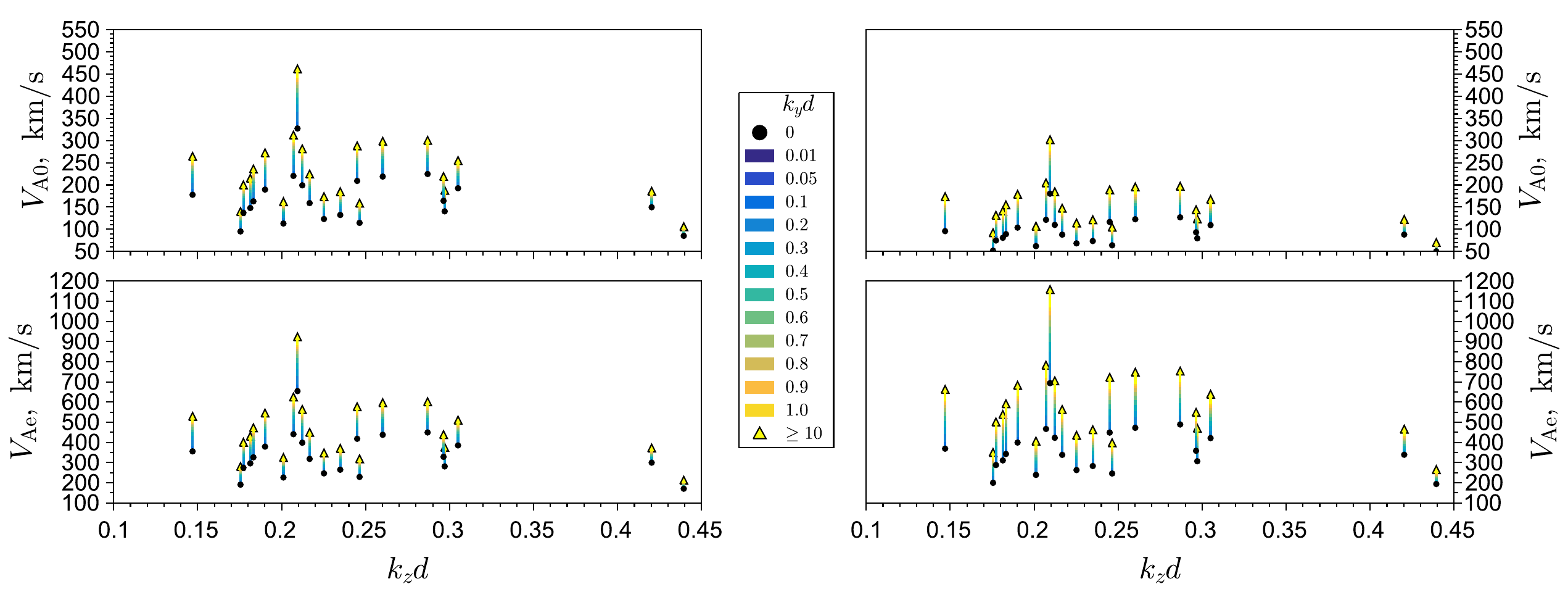}
    \caption{The estimated Alfv\'en speed inside the slab $V_\mathrm{A0}$ (top row) and outside the slab $V_\mathrm{Ae}$ (bottom row) versus $k_z d$ for 22 streamer waves from \citep{2020ApJ...893...78D}. Here $\rho_0 / \rho_e = 4$ and $\beta_\mathrm{e} = 0$. Left and right panels correspond to $\beta_0 = 0$, $V_\mathrm{Ae} = 2 V_\mathrm{A0}$ and $\beta_0 = 2.7$, $V_\mathrm{Ae} = 3.85 V_\mathrm{A0}$ models, respectively. Colour-coded segments for each event show the Alfv\'en speed for different $k_y d$ according to the legend.}
    \label{fig:VAestim}
\end{figure}

The phase speed  calculation, described in Section~\ref{sec:phas}, is normalized to the Alfvén speed inside the slab $\omega/k_z V_\mathrm{A0}$. Thus, the estimated Alfvén speed can be obtained by dividing the observed phase speed by the model phase speed, assuming $\lambda_z / P = \omega/k_z$. We calculated the Alfvén speed inside and outside the slab for both models in the same manner for the set of $k_y d$ we used in Section~\ref{sec:phas} (from 0 to 1), supplemented with $k_y d = 10$ as a proxy for the $k_y \gg k_z$ condition (Figure~\ref{fig:VAestim}).

According to the results presented in Figure~\ref{fig:VAestim}, if wave obliquity is taken into account, the estimated $V_\mathrm{A0}$ can be up to 1.5 times higher than $V_\mathrm{A0}$ for $k_y = 0$ for both models. The values of $V_\mathrm{A0}$ appear to span the range 50--500~\kms. This estimate accounts for both effects of the projection and the decrease in the phase speed of the kink wave caused by obliquity. 

In the zero-$\beta$ model, the estimated $V_\mathrm{Ae}$ follows the $V_\mathrm{A0}$ dependence on $k_y$, while in the $\beta > 1$ model, it might be up to 2 times higher in the case of the oblique propagation than the parallel propagation. Thus, the range of $V_\mathrm{Ae}$ for different $k_y$ spans 150 to 1200~\kms. These values of the external Alfv\'en speed are higher than previously estimated \citep{2011ApJ...728..147C}, which should be attributed to the possible effect of obliquity.
The dependence of the estimated $V_\mathrm{A0}$ and $V_\mathrm{Ae}$ on $k_y$ is not linear with the fastest change in the range $k_y d = 0.1-0.4$ and almost negligible for $k_y d > 10$ and $k_y d < 0.01$.

\begin{figure}
    \centering
    \includegraphics[width=1\linewidth]{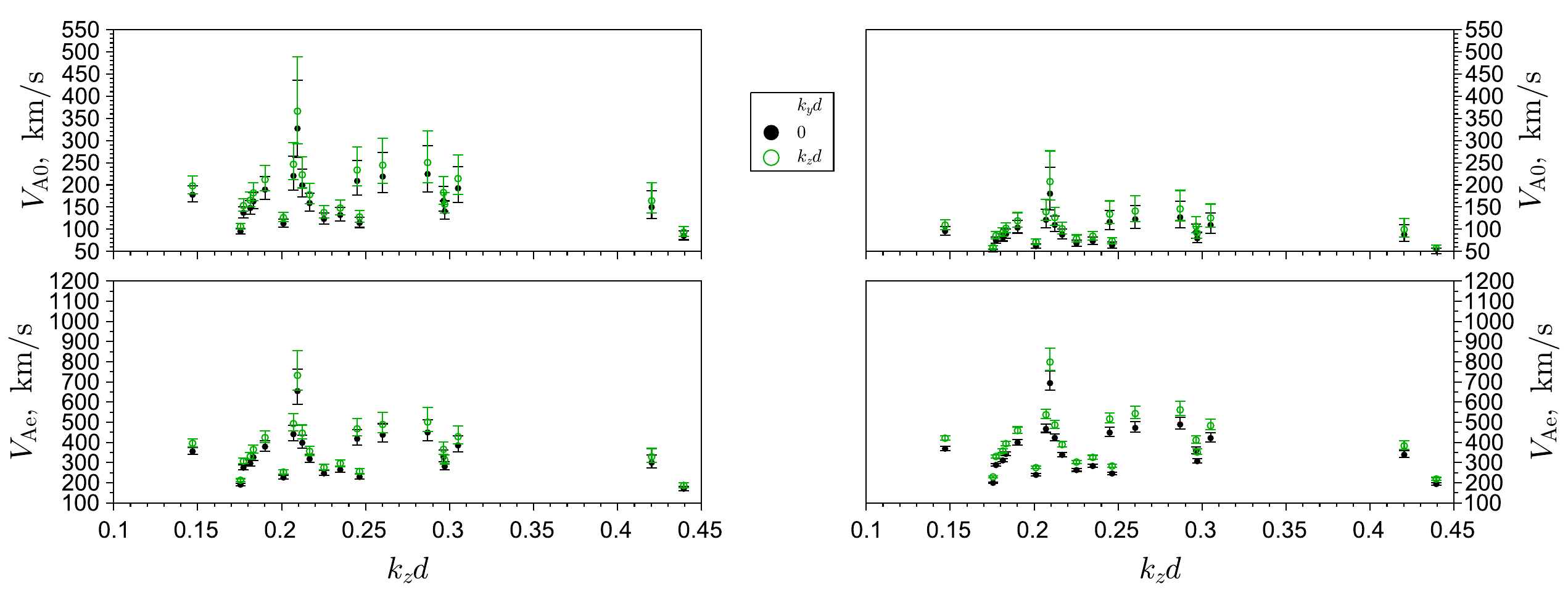}
    \caption{The comparison of the obliquity effect and the period measurement error for the estimated Alfvén speed inside the slab $V_\mathrm{A0}$ (top row) and outside the slab $V_\mathrm{Ae}$ (bottom row) for 22 streamer waves from \citep{2020ApJ...893...78D}. Here $\rho_0 / \rho_e = 4$ and $\beta_\mathrm{e} = 0$. Left and right panels correspond to $\beta_0 = 0$, $V_\mathrm{Ae} = 2 V_\mathrm{A0}$ and $\beta_0 = 2.7$, $V_\mathrm{Ae} = 3.85 V_\mathrm{A0}$ models, respectively. Parallel propagation case is shown in black and oblique propagation under $k_z = k_y$ condition is shown in green. Error bars demonstrate the wave period measurement error.}
    \label{fig:VAerror}
\end{figure}

The uncertainty in the measured wave period, $P$, also affects the estimate of the Alfv\'en speed, and we compared it with the effect of wave propagation obliquity (Figure~\ref{fig:VAerror}). For estimates of $V_\mathrm{A0}$, an obliquity angle of $45^\circ$ ($k_y = k_z$) has a smaller effect than the uncertainty associated with the wave period measurement. In contrast, its impact on the estimates of $V_\mathrm{Ae}$ is considerably more significant. For the $\beta > 1$ model, the effect of oblique propagation pushes the estimated $V_\mathrm{Ae}$ beyond the uncertainty associated with the wave period measurement under the assumption of field-aligned propagation for all observed streamer waves.

The numerical solution of Eq.~\ref{eq:dispeq} demonstrates that the choice of $d$ has little effect on the estimations. The ratio between Alfv\'en speeds of the obliquely propagating wave ($k_y = k_z$) and of the parallel propagating wave ($k_y = 0$) for $d =$ 0.1--0.3~$R_\odot$ varies by about 1\%.

\section{Discussion and Conclusion}
\label{Sec:con}
Coronal streamer kink waves excited by the impact of CMEs are routinely used to estimate the coronal Alfvén speed. These estimates generally assume propagation parallel to the radial magnetic field, although the geometry of CME impacts naturally allows for oblique wave propagation.
In this study, we consider oblique kink waves guided by a plasma slab representing the streamer stalk. The waves propagate at an angle to the equilibrium magnetic field, which is assumed to be directed radially outward from the Sun. At the centre of the slab, there is an infinitely thin current sheet formed by anti-parallel magnetic fields on either side. Effects associated with the spherical geometry, such as the variation of the stalk width with height, solar-wind outflow, magnetic shear, and finite-amplitude effects, are neglected. The slab is surrounded by a cold ($\beta_\mathrm{e} = 0$) plasma. Inside the slab, cases of both cold and warm plasma are considered, with $\beta_0 = 0$ and $\beta_0 > 1$, respectively. 
Special attention is paid to the effect of wave obliquity that is quantified by the ratio of the parallel and perpendicular components of the wave vector, in the plane of the slab, on the seismological estimation of the Alfvén speed in and around the streamer stalk.

According to \citet{1982SoPh...76..239E}, in the case of parallel propagation, the phase speed of kink waves guided by such a slab has a value below the external Alfv\'en speed. In the case of zero-$\beta$ plasma inside the slab, the lower limit of the parallel phase speed is the internal Alfv\'en speed. For internal $\beta > 1$, the lower limit is the internal sound speed. The phase speed increases with an increase in the wavelength. 
The value of the plasma parameter $\beta_0$ inside the slab has a certain effect on the radial phase speed. For waves with $k_zd=1$, the increase in $\beta_0$ from 0 to 2.7 causes a smooth increase in the difference between the radial phase speed and the internal Alfv\'en speed, divided by the difference of the external and internal Alfv\'en speeds from about 0.25 to about 0.33 for a wide range of $k_yd$, from 0 to 1. An interesting special case is $\beta_0 \approx 1$. It should be considered in a generalisation of this model that accounts for the variation of the streamer-stalk parameters with distance from the Sun and investigates the behaviour of kink waves as they pass through the $V_\mathrm{A0}= V_\mathrm{s0}$ layer.

Furthermore, the phase speed of obliquely propagating waves decreases with increasing angle between the equilibrium magnetic field and the wave vector, i.e. the wave obliquity. The phase speed approaches zero in the case of perpendicular propagation. This behaviour is consistent with the findings of \citet{2026A&A...707A.273L}.    

The radial component of the phase velocity, i.e., parallel to the direction of the equilibrium field, decreases with increasing radial component of the wave vector. 
For large radial wavelengths, the radial phase speed decreases from the external Alfv\'en speed value, which corresponds to the limit of parallel propagation, considered by \citet{1982SoPh...76..239E}, with increasing propagation obliquity.
The difference between the radial phase speeds in the case of parallel and oblique propagations decreases with the decrease in the radial wavelength, becoming negligible for the short wavelength limit. 

Parallel kink waves are of the body nature. Body waves are characterised by the highest perturbation amplitude in the centre of the slab, where the current sheet is situated. The increase in the obliquity of the wave propagation gradually spreads the perturbation perpendicularly out of the current sheet, approaching the surface mode regime. In the transition from the body to the surface mode, the perturbation has the same amplitude inside the slab. In the surface regime, the perturbation has a maximum value near the slab boundary, gradually decreasing near the current sheet with the increase in obliquity. 
This effect occurs in both small and large values of $\beta$ inside the streamer stalk.
Similarly, in both cases, the extent of the perturbation outside the slab decreases with the increase in obliquity. 

In the external medium, the amplitude of guided modes decreases exponentially. 
For a given wave energy, the localisation distance of the mode outside the slab determines its amplitude. For wavelengths comparable to the width of the slab, the perturbation amplitude decreases by a factor of $e$ over a distance comparable to the localisation distance. The localisation distance is found to decrease with increasing wave obliquity. Consequently, for a given energy deposition by a CME impact, more oblique modes are expected to have larger amplitudes. This effect is also strengthened by the gradual transition from a body to surface mode structure with the increase in the wave obliquity, which makes the perturbation of the slab boundaries more pronounced.

The theoretical results described above have been applied to the seismological estimation of the Alfv\'en speed in streamers using kink waves detected by \citet{2020ApJ...893...78D}. In contrast to the estimates presented by \citet{2011ApJ...728..147C,2011SoPh..272..119F,2016MNRAS.463.1409S}, we allow for oblique wave propagation, which reduces the phase speed projected onto the plane of the sky. This reduction is caused both by the conventional projection effect and by the intrinsic decrease in the phase speed with increasing propagation obliquity. The inferred Alfv\'en speed spans the range 50--500~\kms. If these effects are neglected, estimates based on the assumption of radial propagation parallel to the equilibrium magnetic field may underestimate the Alfv\'en speed by up to a factor of 1.5. This discrepancy arises only if the waves propagate obliquely, a possibility that cannot be ruled out without true stereoscopic observations.

In a follow-up study it is of interest to account for the effect of the radial flow, i.e., the nascent solar wind, on the seismological inversions of streamer parameters, developing the formalism designed in \citet{1995SoPh..159..213N}. Likewise, the variation of the model parameters with the distance of the Sun, especially the behaviour of the waves at the distance where $\beta_0 = 1$ may be of interest.

Furthermore, in our study, we assumed that the current sheet surrounded by the plasma slab is infinitesimally thin. This assumption is rather standard in the theoretical modelling of coronal streamers \citep[e.g.,][]{2003JGRA..108.1377H} and, in particular, streamer waves \citep[e.g.,][]{2010ApJ...714..644C, 2011SoPh..272..119F, 2011ApJ...728..147C, 2016MNRAS.463.1409S, 2022MNRAS.515.4055G}. This assumption is motivated by the expectation that the intrinsic thickness of a current sheet in a weakly collisional plasma is controlled by kinetic plasma scales, in particular the ion inertial length and the proton gyroradius \citep[e.g.,][]{2009NPGeo..16..443S}, and therefore is much smaller than the characteristic width of the streamer stalk. Since kink waves are collective perturbations of the slab, we expect a narrow, localised transverse non-uniformity associated with the current sheet not to modify their global properties substantially. However, a rigorous demonstration of this expectation would require a dedicated study.

\section*{Acknowledgements}

This study is supported by the National Natural Science Foundation of China (NSFC; grant No. 12473050), the Guangdong Natural Science Fund for Distinguished Young Scholars (grant No. 2023B1515020049), the Shenzhen Science and Technology Project (grant No. JCYJ20240813104805008), and the Science and Technology Program of Guangdong Province (grant No. 2025B1212050001). V.M.N. is supported by ERC grant 101201424 (ACDCSUN) and the BK21 FOUR programme through the National Research Foundation of Korea (NRF) under the Ministry of Education (MoE) (Kyung Hee University, Human Education Team for the Next Generation of Space Exploration), and the Global-Learning \& Academic research institution for Master’s/PhD students, the Postdocs (G-LAMP) Program of the NRF grant funded by the MoE (RS-2025-25442355).

\bibliography{streamer}{}
\bibliographystyle{aasjournalv7}


\end{document}